# Composition-Dependent Properties of $Ce_xLa_{0.95-x}Tb_{0.05}F_3$ Nanopowders Tailored for X-Ray Photodynamic Therapy and Cathodoluminescence Imaging

X. Lytvynenko[1*], M. Urbanová[1], O. Lalinský[2], V. Vojta[2], J. Bárta[1,3], L. Prouzová Procházková[1,3], V. Čuba[1]
[1]*Czech Technical University in Prague, Faculty of Nuclear Sciences and Physical Engineering, Břehová 7, 115 19 Prague, Czech Republic*
[2]*Institute of Scientific Instruments, Czech Academy of Sciences, Královopolská 147, 612 00 Brno, Czech Republic*
[3]*Institute of Physics, Czech Academy of Sciences, Cukrovarnická 10, 162 53 Prague, Czech Republic*
**Corresponding author: popovkse@fjfi.cvut.cz*

**Abstract**

This study investigates the synthesis and luminescence behavior of $Ce_xLa_{0.95-x}Tb_{0.05}F_3$ nanoparticles with varying $Ce^{3+}$ content. The materials were prepared via sol-gel chemical route and thermally annealed to improve crystallinity and reduce defects. Phase composition and structural parameters were examined by X-ray diffraction (XRD), while elemental composition was determined by X-ray fluorescence (XRF). Cathodoluminescence (CL) intensity mapping was used to evaluate emission uniformity and monitor the degradation of luminescence under electron beam exposure. Photoluminescence (PL) and radioluminescence (RL) spectroscopy confirmed energy transfer from $Ce^{3+}$ to $Tb^{3+}$ ions. Luminescence intensities were found to depend strongly on both Ce content and thermal treatment. Based on our measurements, a suitable $Ce^{3+}$ concentration was identified. The results contribute to the understanding of defect-related quenching mechanisms and are relevant for the design of rare-earth-based luminescent nanomaterials for biomedical applications.

## Introduction

Rare-earth-doped lanthanum fluoride nanoparticles have attracted considerable attention in recent years due to their excellent optical properties, good chemical stability, and low phonon energy, which make them prospective candidates for various photonic and biomedical applications (Jacobsohn et al., 2010; Runovski and Lis, 2014; Pudovkin et al., 2019). Cerium-doped lanthanum fluoride ($LaF_3$:$Ce^{3+}$) nanoparticles represent a particularly interesting class of materials owing to the 4f–5d transitions of $Ce^{3+}$ ions, which result in a fast, broadband emission in the UV region (Rodnyi et al., 1995; Dorokhina et al., 2023). The $LaF_3$ host matrix is known for its low vibrational energy and high transparency (Weber, 1967; Hong et al., 2015; Cui et al, 2016), making it an efficient platform for studies of energy transfer processes and luminescence tuning. Chemical composition and nanostructure of $LaF_3$ allow fine-tuning of its optical behavior through controlled doping and synthesis methods (Tian et al., 2009; Gao et al., 2014; Ladol, 2022). Additionally, the co-doping of $Ce^{3+}$ and $Tb^{3+}$ ions can facilitate energy transfer mechanisms that enhance visible emission, further broadening their application potential (Kasturi, 2018).

In parallel, X-ray photodynamic therapy (XPDT) has emerged as a promising modality for cancer treatment, particularly in case of deep-seated tumors. XPDT relies on scintillating nanoparticles that absorb soft ionizing radiation (X-rays) and convert it into UV/visible light, which subsequently activates nearby photosensitizers to produce cytotoxic reactive oxygen species (ROS) (Chuang et al., 2020). For this reason, the design of highly luminescent, radiation-responsive nanoscintillators such as Ce- and Tb-doped $LaF_3$ is of growing interest (Popovich et al., 2016; Dorokhina et al., 2023). Their ability to efficiently convert high-energy photons into biologically relevant optical output is crucial for therapeutic efficacy.

Another important technique that could benefit from these nanoparticles is cathodoluminescence (CL) bioimaging (Furukawa et al., 2015; Bischak et al., 2015; Keevend et al., 2020), which involves the excitation of luminescent materials with electron beams. CL imaging provides nanoscale spatial resolution, enabling detailed visualization of cellular structures and the localization of specific nanoparticles. However, the stability of luminescent properties under electron beam exposure is a key factor for their performance in bioimaging applications.

Previous studies have shown that calcination enhances luminescence by improving crystallinity and reducing non-radiative traps, but a prolonged or excessive annealing often leads to grain growth, dopant redistribution, or concentration quenching, ultimately reducing luminescence efficiency (Rao Bandi et al., 2009; Altunal et al., 2019; El Desouky et al., 2020). In our recent work on $Ce_xLa_{1-x}F_3$ nanoparticles (Lytvynenko et al., 2025), we identified an annealing temperature window that ensures a favorable compromise between luminescence and particle size. The present study extends this approach to Ce and Tb co-doped $LaF_3$, where the crystallinity, emission intensity, and nanoparticle size, properties critical for XPDT and CL applications, need to be balanced.

This work is focused on the synthesis of a series of $Ce_xLa_{0.95-x}Tb_{0.05}F_3$ nanoparticles via a sol-gel route, enabling precise control over composition and morphology. A systematic study was conducted to investigate the effect of $Ce^{3+}$ concentration on the structural, morphological, and luminescent properties of the resulting powders. This work also aims to investigate the energy transfer from $Ce^{3+}$ to $Tb^{3+}$, where the radioluminescence intensity is influenced not only by the transfer efficiency, but also by the powder nature of the samples. Special attention was paid to the stability of CL emission, a critical factor for their applicability in CL imaging and XPDT platforms. The relationship between $Ce^{3+}$ content, crystal structure, and luminescent performance is discussed with respect to optimizing these materials for multimodal theranostic use.

## Experimental

*Materials and methods:*

All $Ce_xLa_{0.95-x}Tb_{0.05}F_3$ samples were prepared using a sol-gel method under identical conditions, with ethylene glycol monomethyl ether (EGME; ≥99%, Penta chemicals) used as the solvent. The exact molar amounts of $Ce(NO_3)_3{\cdot}6H_2O$ (≥99%, Sigma Aldrich), $La(NO_3)_3{\cdot}6H_2O$ (99.999%, Sigma Aldrich), and $Tb(NO_3)_3{\cdot}6H_2O$ (99.999%, Sigma Aldrich) used for the synthesis are summarized in Table 1.

**Table 1:** Quantities of nitrate precursors used for the synthesis. The dopant concentrations are given as nominal values.

| **Sample** | $n_{La(NO_3)_3}$ $[mmol]$ | $n_{Ce(NO_3)_3}$ $[mmol]$ | $n_{Tb(NO_3)_3}$ $[mmol]$ |
|---|---|---|---|
| $La_{0.95}Tb_{0.05}F_3$ | 4.845 | - | 0.260 |
| $Ce_{0.15}La_{0.80}Tb_{0.05}F_3$ | 4.083 | 0.767 | |
| $Ce_{0.30}La_{0.65}Tb_{0.05}F_3$ | 3.321 | 1.534 | |
| $Ce_{0.475}La_{0.475}Tb_{0.05}F_3$ | 2.423 | 2.425 | |
| $Ce_{0.65}La_{0.30}Tb_{0.05}F_3$ | 1.531 | 3.316 | |
| $Ce_{0.75}La_{0.20}Tb_{0.05}F_3$ | 1.025 | 3.832 | |
| $Ce_{0.95}Tb_{0.05}F_3$ | - | 4.848 | |

The nitrate precursors were transferred into a 150 mL glass beaker, followed by the addition of 50 mL of EGME. The mixture was stirred at 50 °C on a magnetic stirrer until complete dissolution was achieved. Simultaneously, 15.3 mmol of $NH_4F$ (≥98%, Penta chemicals) was dissolved in 100 mL of EGME at 75 °C in a quartz or fluoropolymer flask equipped with a reflux condenser. During heating, additional 70 mL of EGME was added to ensure complete dissolution. The solution containing the dissolved nitrates was then slowly poured into the same flask, and the reaction mixture was stirred and maintained at 60 °C for an additional hour. During this stage, the formation of a white gel was observed.

After cooling the mixture to room temperature, the gel was centrifuged (Biofuge Stratos, Sorvall; RCF = 4863 g, 5 min) at 23 °C and washed four times with ethanol to remove residual solvent and unreacted species. The final product was left to dry in air at 40 °C overnight. The samples were annealed for 30 minutes in Labsys Evo thermal analyzer (Setaram instrumentation, France) at 600 °C in an $Ar/H_2$ atmosphere to prevent their oxidation.

Subsequent sample preparation procedures were carried out for studies using a scanning electron microscope (SEM) and CL. The powder was suspended in 100% acetone (VWR) to a concentration of 0.01 mg/µL and dispersed by brief vortexing and ultrasonication. After a short centrifugation step (MiniSpin plus, Eppendorf; RCF = 1073 g, 30 s), 85% of the supernatant was transferred to a new tube. Aliquots (3 × 10 µL) were then deposited onto carbon tape-coated SEM stubs, ensuring uniform spreading without air bubbles. Samples were air-dried at room temperature for 30 minutes prior to analysis.

*Samples characterization:*

X-ray diffraction (XRD) patterns were obtained using Rigaku MiniFlex 600 diffractometer fitted with a Cu X-ray tube ($\lambda_{K\alpha1,2}$ = 0.15418 nm). The measurements were carried out under operating conditions of 40 kV and 15 mA. Data were collected in continuous scanning mode over a 2θ range from 20° to 80°, with a scanning rate of 2° per minute, data collection interval 0.02°. Phase identification was performed using the PDXL2 software in conjunction with the ICDD PDF-2 database (version 2013), and the results were compared to the reference pattern of $LaF_3$ (#32-0483). Crystallite sizes were calculated using the Halder–Wagner linearization (Halder and Wagner, 1966), utilizing the integral breadth of all observed diffraction peaks and Scherrer constant K = 1.0747 (Langford and Wilson, 1978).

X-ray fluorescence (XRF) analysis was performed using a Niton XL3t 900 GOLDD analyzer (Thermo Fisher Scientific, USA), equipped with a 2 W silver X-ray tube, a large silicon drift detector, and a Peltier cooling system. Measurements were conducted in three operating modes: main range (50 kV, 40 µA, Al filter, 250 s), low range (20 kV, 100 µA, Cu filter, 500 s) and light range (8 kV, 200 µA, 250 s). The obtained spectral data were evaluated against reference energies from the NIST SRD 128 standard database. Due to its Ag X-ray tube, characteristic lines of Ag (22, 25, and 3 keV) as well as their Compton scattering bands (21 and 23 keV) are present in the spectra.

To evaluate luminescent properties of the prepared samples, both room temperature (RT) photoluminescence (PL) and radioluminescence (RL) spectra were collected. Steady-state PL spectra were recorded on a Fluoromax 4-Plus spectrofluorometer (Horiba) equipped with 150 W high-pressure Xe discharge lamp and a R928P photomultiplier detector. The solid sample holder with an adjustable goniometer was set to 30° in all measurements to prevent the primary beam from being directly reflected into the detector. Measured intensities were corrected for the spectral dependence of detection sensitivity and excitation beam intensity. RL emission spectra were recorded using a tungsten X-ray tube (Seifert) operating at 40 kV and 15 mA as the excitation source. The detection system consisted of a Horiba Jobin Yvon 5000M spectrofluorometer equipped with a single-grating monochromator and a TBX-04 photon-counting detector. Spectral data were corrected for the wavelength-dependent sensitivity of the detection setup. To enable quantitative comparison of steady-state RL intensity across different samples, the measurements were normalized to the maximum intensity of a BGO (bismuth germanate, $Bi_4Ge_3O_{12}$) powder standard measured under identical experimental conditions. The choice of a powdered BGO reference, rather than a single crystal, was made to account for enhanced light scattering effects typical of powder samples and to ensure a more representative comparison.

Helios G4 HP scanning electron microscope (SEM) equipped with immersion mode and Delmic SPARC CL detector was used for topographic imaging and CL intensity mapping. Measurements were performed with a beam energy of 10 keV and a beam current of 270 pA. The dwell time was set to 250 µs for as-prepared samples and 1 ms for annealed ones. The electron beam current values provided by the SEM were calibrated using a reference experiment with a Faraday cup. Due to the very weak CL signal of the as-prepared samples, frame summation of 20 individual images was necessary to obtain satisfactory results. As a result, the total effective dwell time for as-prepared samples was 5 ms.

Beam-induced degradation was investigated by monitoring time-dependent changes in CL intensity during repeated continuous scanning of a 2 × 2 µm area containing a higher concentration of nanoparticles. A dwell time of 50 ns, a beam energy of 10 keV, a beam current of 22.8 pA, and a frame time of 300 µs were used for each scan. A similar value of the current density × time [$C/cm^2$] was used as for CL intensity mapping. Whereas uncoated samples were used for topography and CL intensity mapping, a 50 nm layer of indium tin oxide was sputter-coated onto samples used for degradation studies to eliminate charging effects.

**Results and discussion**

XRD analyses were performed to verify the phase composition of the samples. Fig. 1a shows the diffraction patterns of the as-prepared powders, while Fig. 1b presents those annealed at 600 °C in an $Ar/H_2$ atmosphere. Annealing resulted in sharper diffraction peaks, indicating increased crystallite size (Table 2). A slight peak shift toward higher angles with increasing $Ce^{3+}$ content reflects changes in lattice parameters (see Fig. 1b inset and Tables S1 and S2 in Supplementary Materials). The as-prepared samples featured crystallite sizes ~5 nm, while annealed samples, especially those with higher $Ce^{3+}$ content, exceeded 100 nm; this corresponds to sintering of the particles upon annealing.

**Table 2:** Crystallite sizes of the as-prepared and annealed $Ce_xLa_{0.95-x}Tb_{0.05}F_3$ samples.

| **sample** | **crystallite size [nm]** | |
|---|---|---|
| | **as-prepared** | **annealed** |
| $La_{0.95}Tb_{0.05}F_3$ | 4.4 ± 0.7 | 84 ± 14 |
| $Ce_{0.15}La_{0.8}Tb_{0.05}F_3$ | 4.5 ± 0.4 | 74 ± 12 |
| $Ce_{0.3}La_{0.65}Tb_{0.05}F_3$ | 4.5 ± 0.8 | 84 ± 15 |
| $Ce_{0.475}La_{0.475}Tb_{0.05}F_3$ | 4.1 ± 0.5 | 78 ± 18 |

| $Ce_{0.65}La_{0.3}Tb_{0.05}F_3$ | 5.0 ± 1.1 | 123 ± 12 |
|---|---|---|
| $Ce_{0.75}$ $La_{0.2}Tb_{0.,05}F_3$ | 4.9 ± 1.3 | 118 ± 21 |
| $Ce_{0.95}Tb_{0.05}F_3$ | 5.1 ± 1.2 | 108 ± 13 |

Secondary electron (SE) imaging and CL intensity mapping were performed to evaluate the spatial homogeneity of CL. Results for the annealed $Ce_{0.15}La_{0.8}Tb_{0.05}F_3$ are shown in Fig. 2, while additional SE and CL images of other samples are provided in the Supplementary Materials (Fig. S1 – S3). Fig. 2a,b were acquired simultaneously from the same sample area. The CL image (Fig. 2b) demonstrates that the nanoparticles emit CL with comparable intensity, indicating uniform distribution. High-resolution SE imaging in immersion mode (Fig. 2c) revealed well-defined nanoparticles or small agglomerates, with sizes of 80–100 nm consistent with XRD data (Table 2). Across all the regions investigated, the particles exhibited comparable sizes to those shown in Fig. 2, with a moderately broad distribution. This suggests that the sample morphology is relatively uniform on the microscale, despite the powder nature of the material. To illustrate this, two SE images of the annealed $Ce_{0.15}La_{0.8}Tb_{0.05}F_3$ sample, acquired from different large areas of the sample, are provided in the Supplementary Materials (Fig. S4).

Electron beam-induced degradation of the nanoparticles was studied by monitoring changes in CL intensity over time (Fig. 3). The results indicate significantly lower beam-induced degradation in the annealed samples compared to the non-annealed ones. A transient increase in CL intensity observed for the annealed $La_{0.95}Tb_{0.05}F_3$ sample is likely related to the "bright burn" effect, where progressive filling of deep traps enhances radiative recombination at luminescent centers (Fasoli et al., 2007; Moretti et al., 2014; Moretti et al., 2016). This phenomenon was observed only in the annealed $La_{0.95}Tb_{0.05}F_3$ sample and warrants further investigation.

Lattice parameters of the samples were also determined by XRD and compared to the average size of rare-earth ions in the lattice. Average ionic radii were calculated from Ce, La, Tb content and the tabulated crystal radii for coordination number 9 (Shannon, 1976). Fig. 4 shows that both as-prepared and annealed samples feature unit cell volume that linearly increases with La content. This increase is in agreement with the increased crystal radius of $La^{3+}$ compared to $Ce^{3+}$ and roughly follows the linear fit of the unit cell volumes of all hexagonal / trigonal $\{RE\}F_3$ versus the crystal radius of $\{RE\}^{3+}$. Lattice parameter determination in as-prepared powders was less precise due to the broad and overlapping diffraction peaks as well as their lower intensity.

XRF analysis was used to determine the elemental composition, detect impurities, and verify the Ce, La, and Tb ratios in the $Ce_xLa_{0.95-x}Tb_{0.05}F_3$ prepared samples, fluorine being non-detectable in XRF. In the Main range spectra, impurities were found only in the $Ce_{0.75}La_{0.2}Tb_{0.05}F_3$ sample (Fig. 5a), where Y-$K_\alpha$ (15 keV), probably from the $Ce(NO_3)_3 \cdot 6H_2O$ precursor (purity ≥ 99%), was detected. The Low range spectra exhibited a small 1.5 keV peak of Al, which is an instrumental artefact. The complete list of impurities obtained from the XRF analysis is provided in the Supplementary Material (Table S3). The intensity of overlapping La and Ce L-lines varied with composition: in $Ce_{0.15}La_{0.8}Tb_{0.05}F_3$ (Fig. 5b), the La peaks dominate over Ce peaks, while in $Ce_{0.75}La_{0.2}Tb_{0.05}F_3$ (Fig. 5c) and $Ce_{0.95}Tb_{0.05}F_3$ (Fig. 5d), Ce peaks are stronger than La due to higher Ce content. Light range spectra (Fig. 5e, linear scale) also showed Ce and La bands whose intensity reflects sample composition. Additional XRF spectra are presented in Supplementary Materials (Fig. S5 – S7).

Quantitative analysis was based on Low range XRF spectra. Ce and La significantly overlap, so their contents were determined from the multiple-Gaussian fit of the $L_\iota$-$L_\alpha$ region (3.9-5.0 keV). Tb content was calculated from the ratio of the Tb-$L_\beta$ line (6.98 keV) to the Ce or La signal. Fig. 6a shows a linear relationship between $L_\iota$ and $L_\alpha$ line intensity ratios of Ce and La and their molar ratios, confirming that the actual Ce and La contents most likely match the target compositions; despite higher overlap, the strong $L_\alpha$ lines provide a more precise estimate of Ce/La content. Tb content was verified via intensity ratio vs. molar ratio plots for element pairs Tb/La (Fig. 6b) and Tb/Ce (Fig. 6c). The samples followed linear fits with high $R^2$ in these plots, again better for $L_\alpha$ lines. The $Ce_{0.15}La_{0.8}Tb_{0.05}F_3$ sample was excluded from Fig. 6c due to possible inaccuracies in Ce determination, caused by a weak Ce-$L_\iota$ line fitting. The other samples followed an approximately linear trend. The $Ce_{0.475}La_{0.475}Tb_{0.05}F_3$ sample was excluded from this analysis due to a phase impurity observed in the sample (Supplementary Material, Fig. S8).

RT PL excitation and emission spectra (Fig. 7) confirm the presence of a broad band corresponding to the $Ce^{3+}$ 4f–5d transition in the excitation spectrum of Tb, as well as lines corresponding to $Tb^{3+}$ 4f–4f transitions in the emission spectrum of Ce, indicating energy transfer from Ce to Tb. For $Tb^{3+}$ emission, a 399 nm cutoff filter was used to block the second harmonics of the excitation light. The samples (except $La_{0.95}Tb_{0.05}F_3$, Fig. 7a) show in the $Tb^{3+}$ excitation spectra (Fig. 7b, $\lambda_{em}$ = 543 nm) a $Ce^{3+}$-related excitation band at 300 nm, followed by $Tb^{3+}$ excitation lines up to ~385 nm, confirming $Ce^{3+} \longrightarrow Tb^{3+}$ energy transfer. Emission spectra for $\lambda_{ex}$ = 300 nm ($Ce^{3+}$ 4f–5d) show a broad 5d–4f band at ~325 nm as well as $Tb^{3+}$ emission lines. $Tb^{3+}$ emission features 4f–4f transitions with four main bands ($^5D_4 \longrightarrow {}^7F_{6,5,4,3}$) between 475 - 630 nm, the strongest at ~540 nm. Band positions remained constant across compositions (see Fig. S9 in Supplementary materials), which is typical for 4f–4f emitting lanthanide ions. An unknown PL center (~500 nm broad band) was detected in emission spectra of $Ce_{0.3}La_{0.65}Tb_{0.05}F_3$ (Fig. 7c, grey solid line), which was probably caused by some trace impurities originating from the precursors, or possibly by a minor phase impurity below the XRD detection limit.

RT RL spectra were measured to quantify luminescence intensity and compare emission spectra with PL results. In the spectra of as-prepared samples (Fig. 8a,b), $La_{0.95}Tb_{0.05}F_3$ and $Ce_{0.15}La_{0.8}Tb_{0.05}F_3$ show higher $Tb^{3+}$ emission. Other as-prepared samples show lower RL intensities which we ascribe to concentration quenching with increasing $Ce^{3+}$ content (see below). $Ce^{3+}$ 5d–4f emission appears at ~325 nm and is strongest in $Ce_{0.475}La_{0.475}Tb_{0.05}F_3$ (slightly red-shifted). $Tb^{3+}$ 4f–4f emissions (475 - 630 nm), strongest at ~543 nm, match the results derived from PL spectra. Identical features were observed in the CL spectra as well (see Fig. S10, Supplementary Material).

Luminescence intensities of the annealed samples (Fig. 8c,d) increase when compared to as-prepared samples, most probably due to trap healing during the annealing. The annealed $Ce_{0.15}La_{0.8}Tb_{0.05}F_3$ sample shows the strongest $Ce^{3+}$ and $Tb^{3+}$ emissions. The $Ce^{3+}$ 5d–4f emission band shifts to ~290 nm after annealing (Fig. 8c,d), which is consistent with the single crystal data (Nikl et al., 1995). The as-prepared samples feature a red-shifted Ce emission, which was attributed to defect-perturbed Ce centers in (Lytvynenko et al., 2025). Additional weak $^5D_3 \rightarrow {}^7F_{6,5,4}$ transitions of $Tb^{3+}$ (350–450 nm) are visible as well. Dominating emission from the $^5D_4$ level suggests the presence of a non-radiative cross-relaxation process of the $^5D_3$ level caused by high concentration of $Tb^{3+}$ (5%) (Álvarez-Ramos et al., 2023). The results in Table 3 show that annealing leads to an increase in luminescence intensity by up to 9× and indicate that the optimal $Ce^{3+}$ concentration in co-doped samples exists: at low $Ce^{3+}$ content, the energy transfer slightly enhances $Tb^{3+}$ emission, while at higher $Ce^{3+}$ concentrations the concentration quenching reduces the RL intensity. The observed enhancement of RL intensity after annealing is consistent with previous reports (Rao Bandi et al., 2009; Altunal et al., 2019; El Desouky et al., 2020). At the same time, these studies also emphasize that excessive annealing can degrade performance due to particle growth and dopant redistribution. To avoid this, we employed annealing conditions identified in our earlier work (Lytvynenko et al., 2025), which provided an optimum balance: particles remain relatively small (80–100 nm) while exhibiting up to 9× higher RL intensity, at the optimum $Ce^{3+}$ concentration of 15%.

Luminescence intensity was also plotted against $Ce^{3+}$ concentration (Fig. 9). For as-prepared samples, an approximately exponential decrease was observed with rising Ce concentration, while annealed samples exhibit a roughly linear decrease. This may be related to the increasing efficiency of the energy transfer along the $Ce^{3+}$ sublattice until a trap is encountered (i.e., concentration quenching); at higher Ce concentration, the efficiency is high and more charge carriers are lost to traps.

**Table 3:** Maximum RL intensity of the as-prepared and annealed $Ce_xLa_{0.95-x}Tb_{0.05}F_3$ samples normalized to BGO standard.

| **Sample** | **Maximum intensity of $Tb^{3+}$ [arb.u.]** |
| --- | --- |

| | as-prepared | annealed |
|---|---|---|
| $La_{0.95}Tb_{0.05}F_3$ | 1.12 | 4.45 |
| $Ce_{0.15}La_{0.8}Tb_{0.05}F_3$ | 0.78 | 5.31 |
| $Ce_{0.3}La_{0.65}Tb_{0.05}F_3$ | 0.44 | 3.95 |
| $Ce_{0.475}La_{0.475}Tb_{0.05}F_3$ | 0.25 | 2.21 |
| $Ce_{0.65}La_{0.3}Tb_{0.05}F_3$ | 0.42 | 2.66 |
| $Ce_{0.75}La_{0.2}Tb_{0.05}F_3$ | 0.41 | 3.85 |
| $Ce_{0.95}Tb_{0.05}F_3$ | 0.41 | 1.92 |

**Conclusions:**

This study demonstrates that the structural and luminescent properties of $Ce_xLa_{0.95-x}Tb_{0.05}F_3$ nanoparticles can be efficiently tuned by varying the Ce/La ratio and applying thermal annealing. XRD analysis confirmed enhanced crystallinity and particle growth upon annealing, accompanied by lattice parameter changes related to $Ce^{3+}$ incorporation. XRF confirmed the elemental composition and supported the targeted molar ratios. CL intensity mapping showed homogeneous luminescence from nanoparticles, while PL and RL spectra confirmed efficient energy transfer from $Ce^{3+}$ to $Tb^{3+}$. For the $La_{0.95}Tb_{0.05}F_3$ annealed sample, an increase in CL intensity was observed, which was preliminarily attributed to the "bright burn" effect and the presence of a moderate amount of deep traps. The nature of this response of the sample warrants further investigation. In the as-prepared samples, however, a much higher degradation of CL intensity than in annealed samples was observed. We attribute the increase in radiation stability to healing of defects, including the surface defects.

Annealing led to a substantial increase in the luminescence intensity (up to 9×), as well as caused a blue shift of the $Ce^{3+}$ emission band, attributed to healing of defects and improved local environments of Ce ions. The $Ce^{3+}$ content strongly influenced luminescence efficiency, with higher concentrations showing quenching effects due to trap-related energy loss.

These findings are directly relevant for applications in both CL imaging, where low electron beam-induced degradation and homogeneous emission are essential for reliable image contrast, and in XPDT, which benefits from efficient RL and controlled energy transfer within nanoscale scintillators. Among the studied compositions, $La_{0.95}Tb_{0.05}F_3$ and $Ce_{0.15}La_{0.80}Tb_{0.05}F_3$ exhibited the most favorable combination of low beam-induced degradation and strong, composition-tunable luminescence, highlighting their potential for use in both advanced CL imaging and XPDT.

**Acknowledgment:**

The work is supported by OP JAC financed by ESIF and the MEYS (Project No. LASCIMAT - CZ.02.01.01/00/23_020/0008525), by the Czech Academy of Sciences (Project No. RVO:68081731), by the Czech Science Foundation under project No. GA23-05615S, and by the Grant Agency of the Czech Technical University in Prague under Grant SGS23/189/OHK4/3T/14.

***Figure captions:***

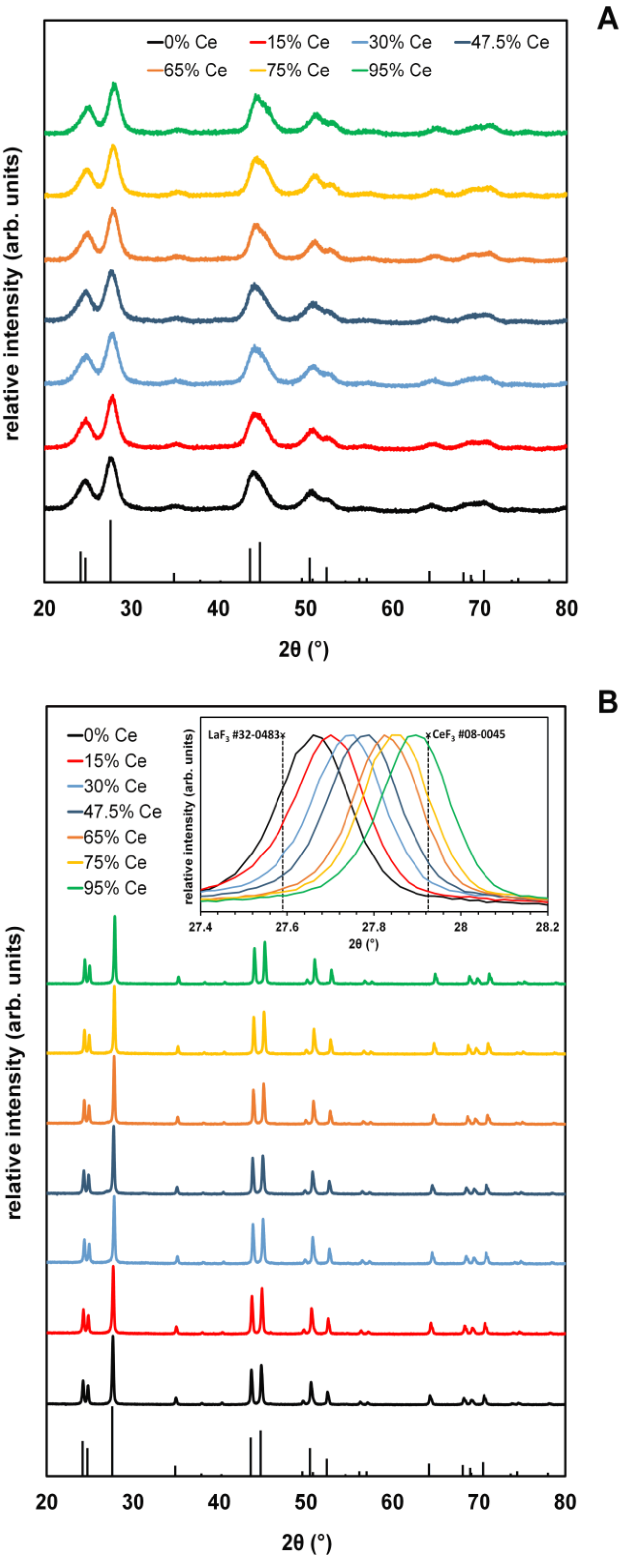

**Fig. 1:** X-ray diffraction patterns of the as-prepared (a) $Ce_xLa_{0.95-x}Tb_{0.05}F_3$ samples and the samples annealed at 600 °C in $Ar/H_2$ (b). The diffraction peak in the region of 27.4 - 28.1° 2θ compared with both reference patterns of $LaF_3$ and $CeF_3$ from the ICDD PDF-2 database is in (b) inset.

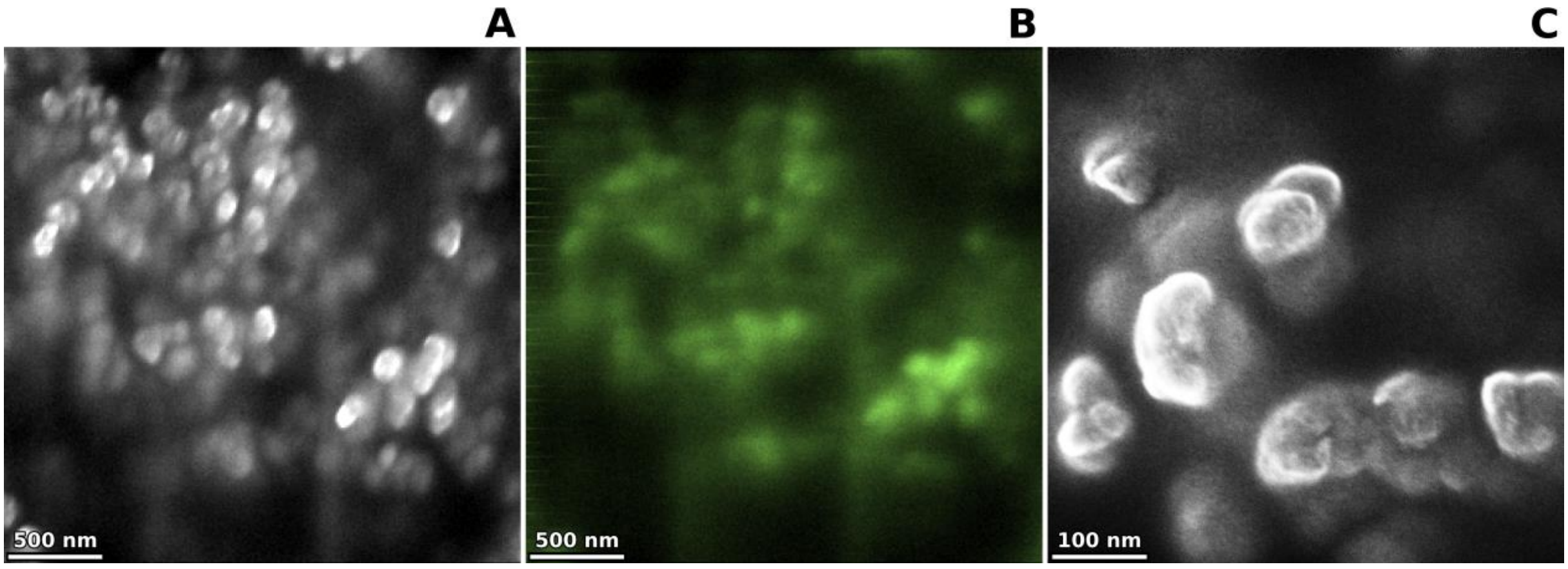

**Fig. 2:** Topography and CL intensity mapping of the annealed $Ce_{0.15}La_{0.8}Tb_{0.05}F_3$. SE image of the sample (a). Corresponding CL intensity map (b). High-resolution SE image in immersion mode (c).

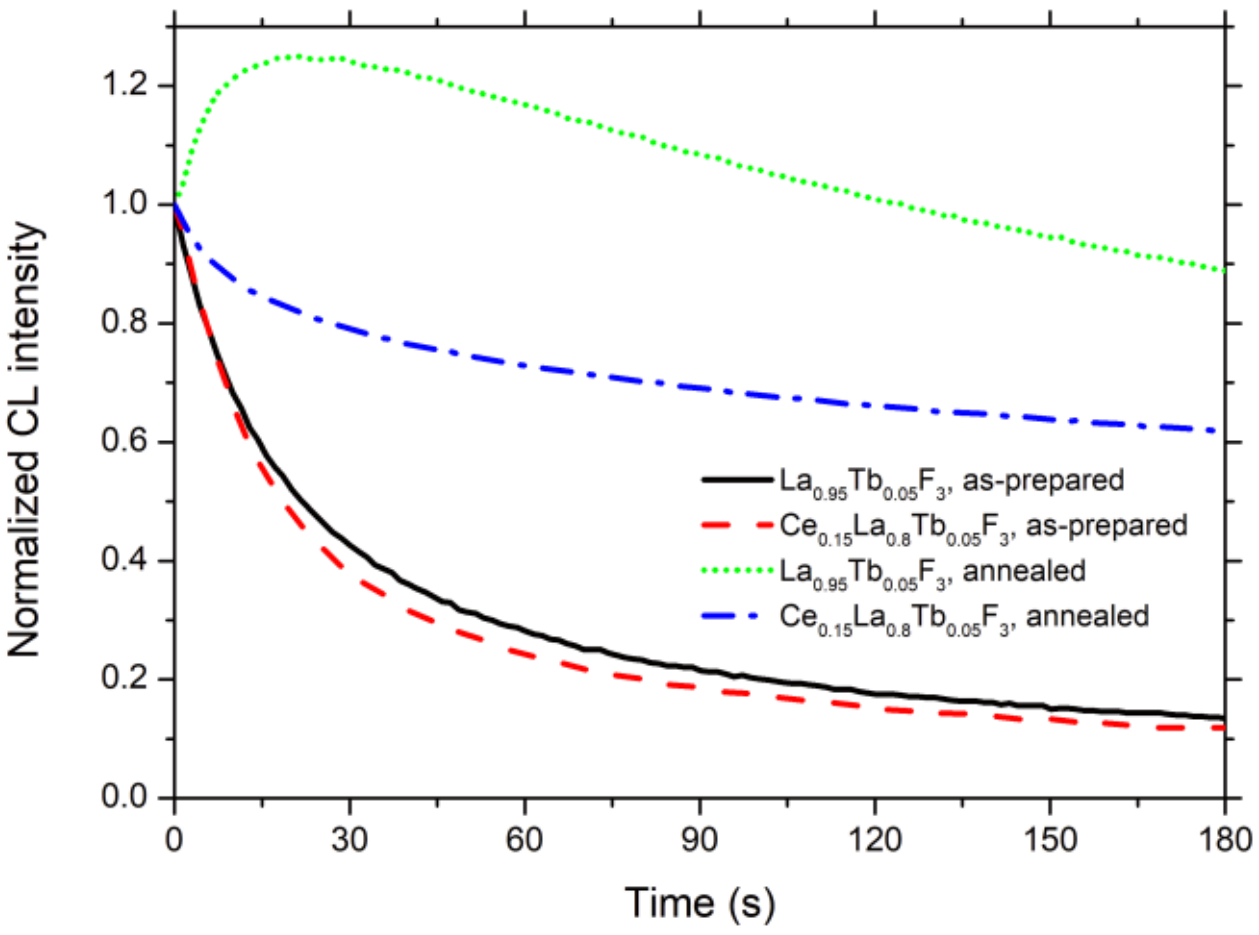

**Fig. 3:** CL intensity as a function of e-beam irradiation time. CL intensities were normalized to the intensity at time = 0.

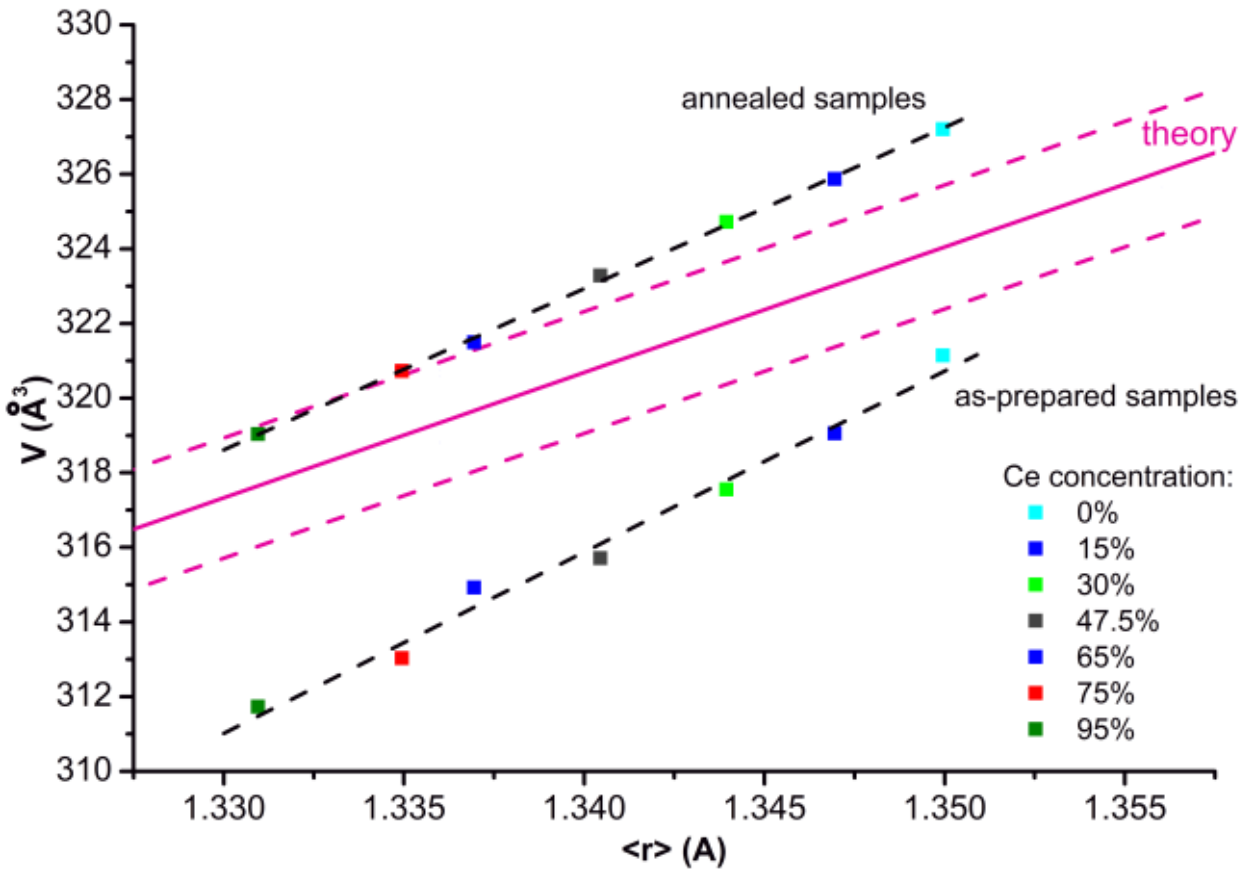

**Fig. 4:** Dependence of the unit cell volume on the average crystal radius for the as-prepared and annealed $Ce_xLa_{0.95-x}Tb_{0.05}F_3$ samples compared to the fit of all $\{RE\}F_3$ ("theory").

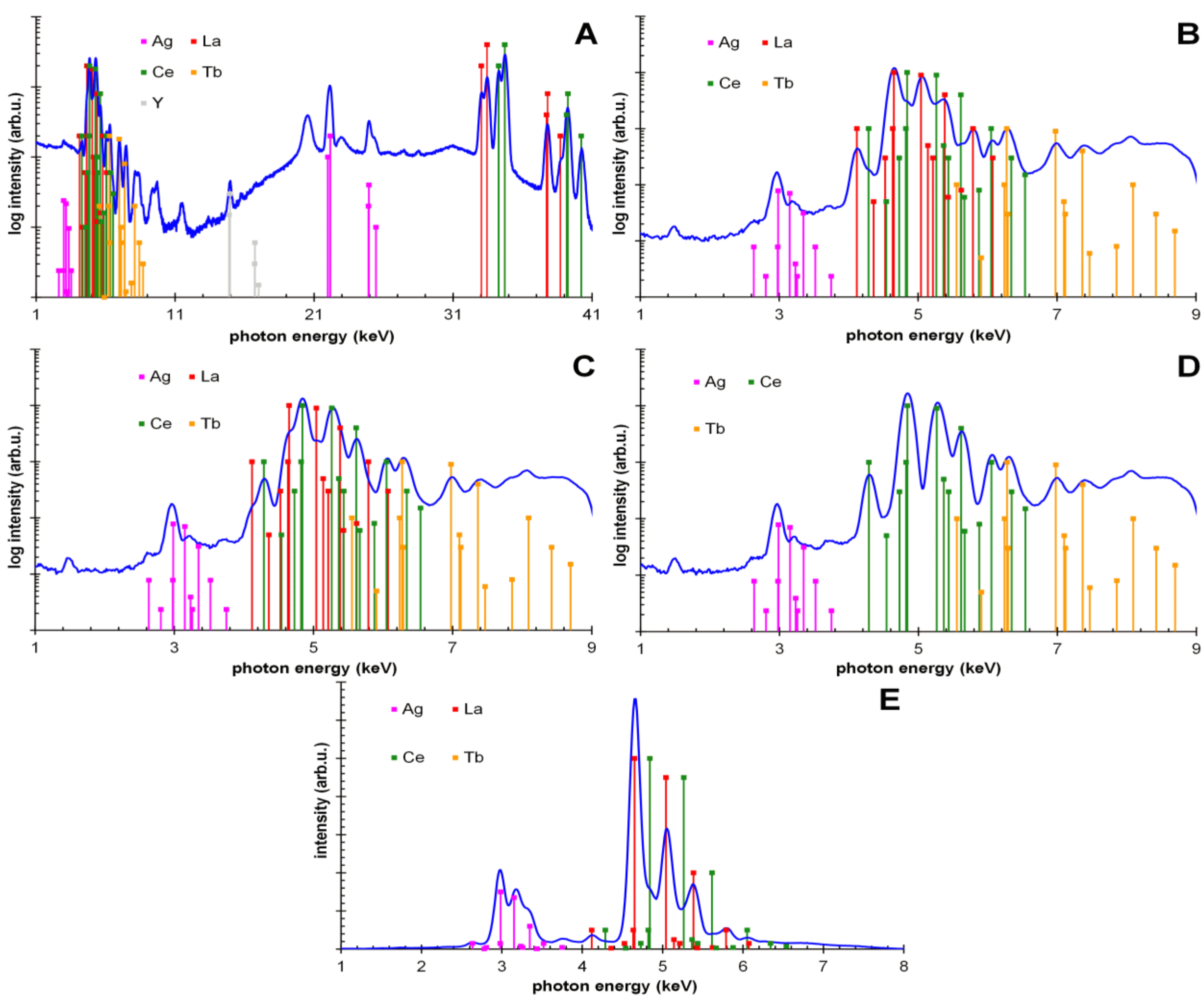

**Fig. 5:** XRF Main range spectrum of the $Ce_{0.75}La_{0.2}Tb_{0.05}F_3$ sample (a), Low range spectra of the $Ce_{0.15}La_{0.8}Tb_{0.05}F_3$ (b), $Ce_{0.75}La_{0.2}Tb_{0.05}F_3$ (c) , $Ce_{0.95}Tb_{0.05}F_3$ (d) samples, and Light range spectrum of the $Ce_{0.15}La_{0.8}Tb_{0.05}F_3$ sample (e).

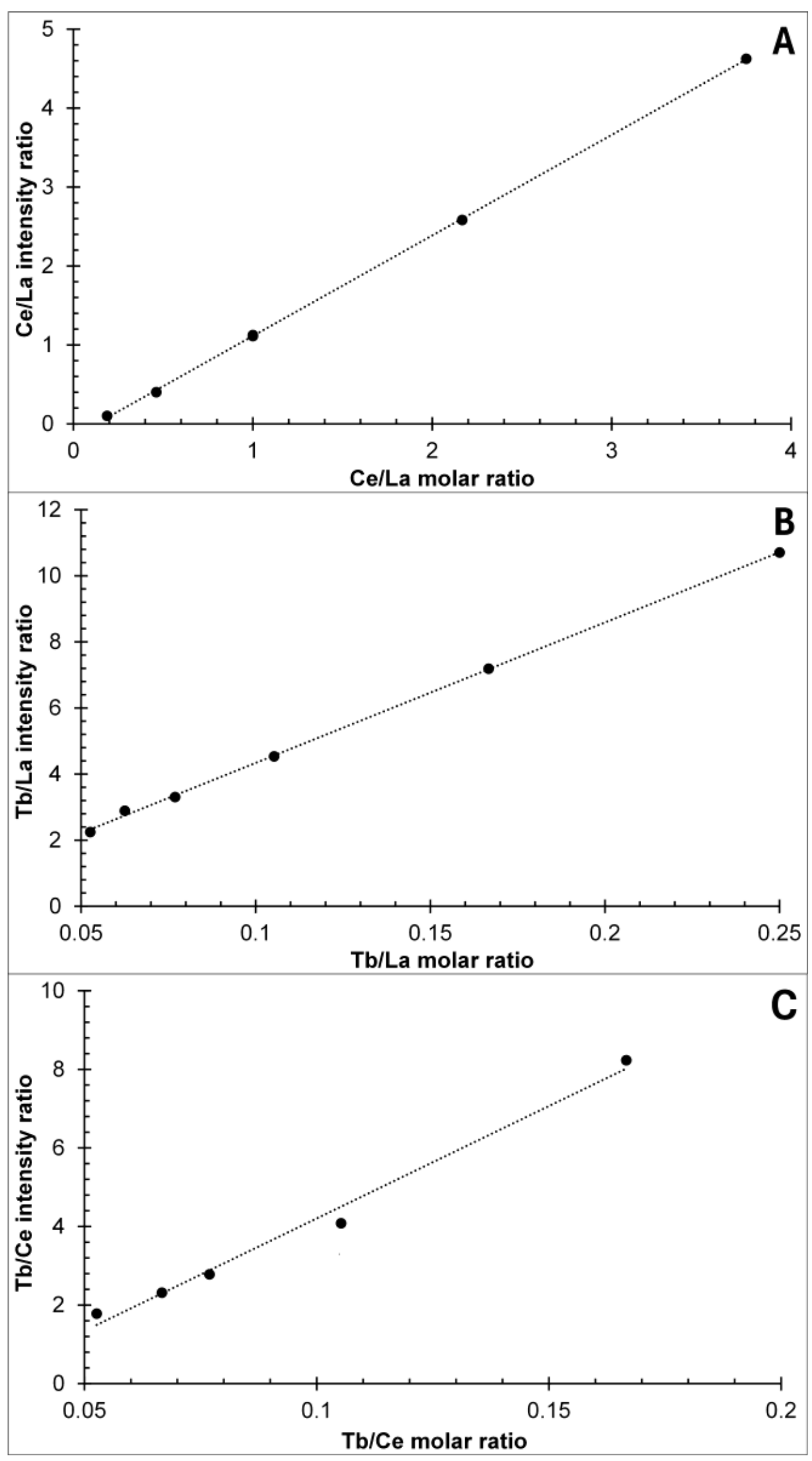

**Fig. 6:** Linear fits of the dependence of the XRF intensity ratio on Ce-$L_{\alpha}$/La-$L_{\alpha}$ (a), Tb-$L_{\beta}$/La-$L_{\iota}$ (b) and Tb-$L_{\beta}$/Ce-$L_{\iota}$ (c) molar ratios.

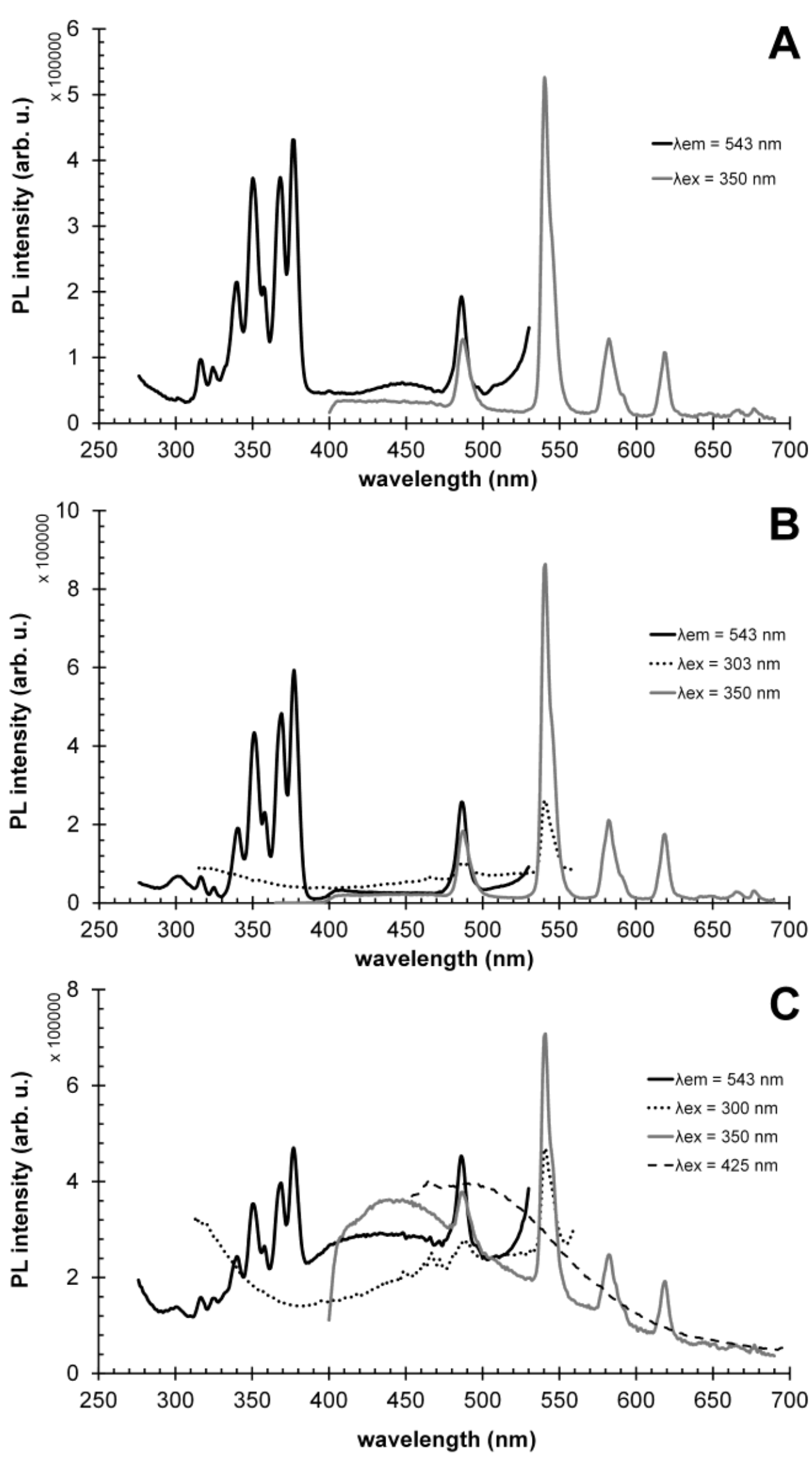

**Fig. 7:** RT PL excitation and emission spectra of the as-prepared $La_{0.95}Tb_{0.05}F_3$ (a), $Ce_{0.15}La_{0.8}Tb_{0.05}F_3$ (b) and $Ce_{0.3}La_{0.65}Tb_{0.05}F_3$ (c) samples.

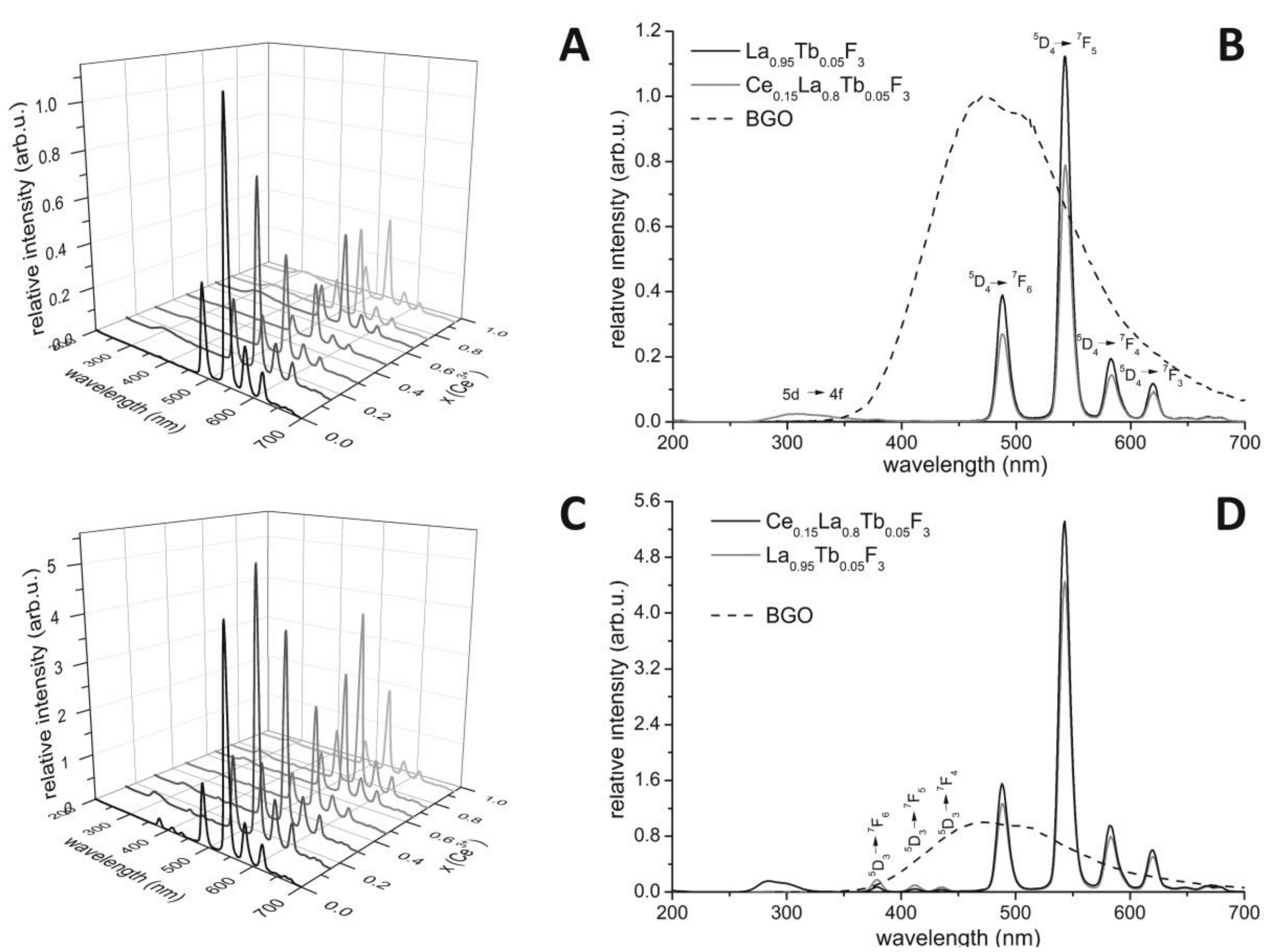

**Fig. 8:** RT RL spectra of the as-prepared (a,b) and annealed (c,d) $Ce_xLa_{0.95-x}Tb_{0.05}F_3$ samples.

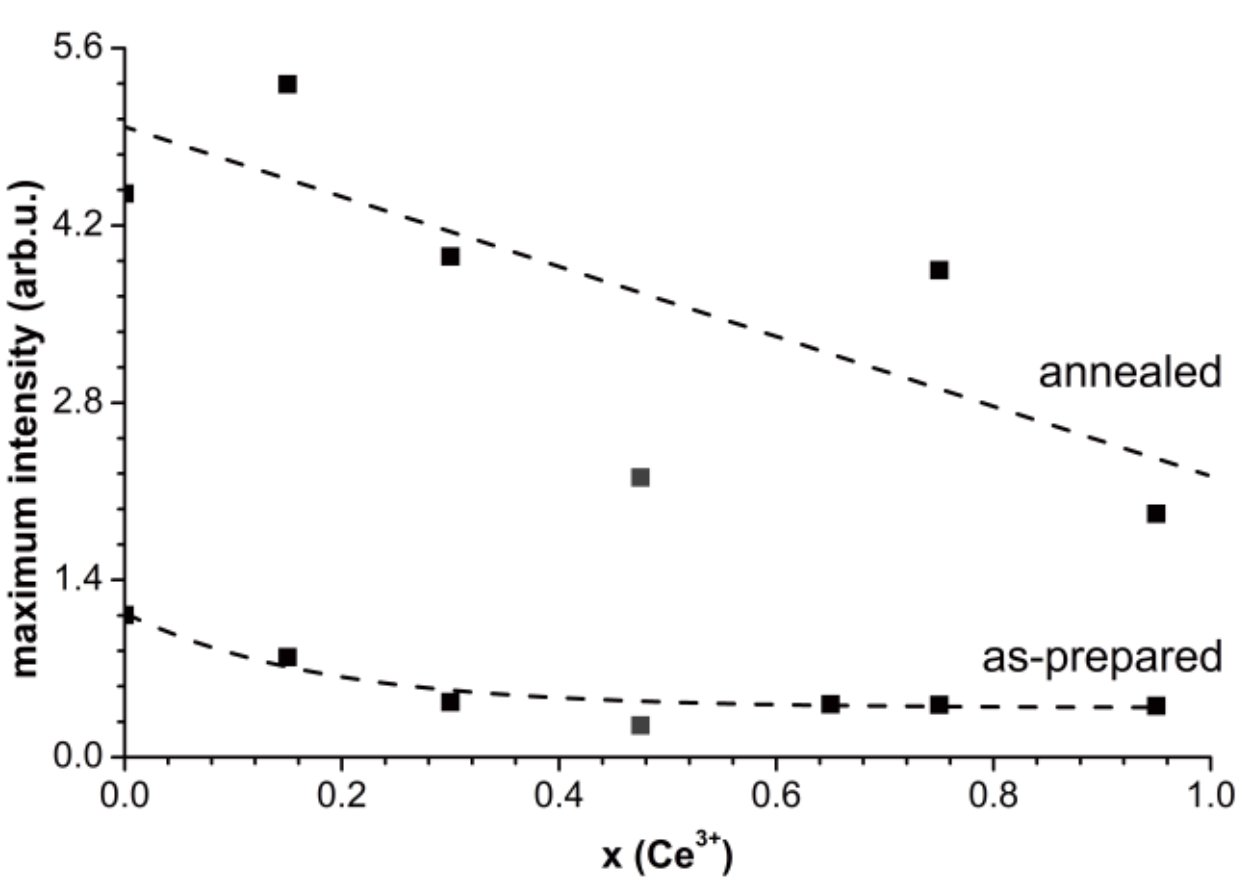

**Fig. 9:** Dependence of the maximum RL intensity of $Ce_xLa_{0.95-x}Tb_{0.05}F_3$ samples on the $Ce^{3+}$ content.